# Stationary and quasi-stationary light pulse in three-level cold atomic system

S.A. Moiseev[1,2] *, A.I. Sidorova[2] and B.S. Ham[3]

[1] E.K. Zavoisky Kazan Physical-Technical Institute, Russian Academy of Science, Kazan, Russia
[2] Kazan Federal University, Kazan, Russia
[3] Gwangju Institute of Science and Technology, Gwangju, Republic of Korea

* E-mail: samoi@yandex.ru

**Abstract**

We have studied stationary and quasi-stationary signal light pulses in cold Λ-type atomic media driven by counterpropagating control laser fields at the condition of electromagnetically induced transparency. By deriving a dispersion relation we present spectral and temporal properties of the signal light pulse and a significant influence of atomic decoherence on the coupled stationary light pulses for spatial splitting. Finally we discuss quasi-stationary light pulse evolution characterized by frozen spatial spreading for a robust coherent control of slow light pulses.

**Introduction**

Quantum coherence control and deterministic manipulation of light pulses have become a topical issue in optics and optical quantum information science. Herein, using electromagnetically induced transparency (EIT) effect [1] is one of the promising tools for the coherent control of weak light fields even in a single photon level [2,3]. This technique is intensively elaborated for various applications of EIT-based quantum memories [4,5] and generation of nonclassical light fields [6]. EIT-based quantum coherence control of weak lights is a strong candidate for ultralow power nonlinear optics [7] applicable to quantum information science such as implementation of two-qubit gates in the optical quantum computing [8-14]. Using stationary light pulse (SLP) concept proposed recently under the EIT condition [15-20] seems to be especially prospective for this purpose due to prolonged lifetime of the slow light pulses in the atomic media. The SLP coherent control with two counterpropagating light fields has been initially demonstrated in a hot atomic gas [15] and in solids [21]. Later this basic scheme was studied for the cold atomic systems in a series of works [22-31]. Herein, the two counterpropagating control light fields lead to the excitation of many spatial gratings and long-lived atomic coherences [22, 23, 25]. The coherence gratings can results in the signal light dynamics of splitting the input light pulse into two counter-propagating light pulses [22, 24-26, 29]. Recent numerical studies [27-29] of nonadiabatic effects in SLP dynamics have revealed strong negative influence of the multiply excited coherence gratings on the SLP [30-31]. However the spectral and temporal properties of SLP remains insufficiently elaborated for SLP outside of the adiabatic coherent control [24, 26]. The problem of light-atom dynamics is due to complicated nonlinear Bragg-scattering of the light fields on the three-level atomic system driven by intensive counterpropagating laser fields. Various atomic relaxation processes have brought in so many important issues of SLP but are still open and require further investigations for spectral properties, time-space dynamics as well as influence of arbitrary intensities of the control laser fields.

In this work, developing coupled Maxwell–Bloch equations for slowly varying amplitudes [22-24,26] with nonadiabatic evolution, we have derived the dispersion relation of stationary light field modes in the cold atomic media. Using a simple dispersion relation we have revealed basic spectral properties of SLP and quasi-stationary light pulses. Herein, we have found narrowed spectral domain providing realization of SLP, and analyzed the temporal properties of the stationary and quasi-stationary signal light pulses. Then, a drastic influence of the atomic decoherence and control laser fields on the basic properties of SLP are demonstrated. In particular, we have observed a new regime of the quasi-stationary light pulse evolution characterized by frozen spatial spreading in the presence of the control laser fields with appropriate intensities. Finally, we describe a simple picture for generation of stationary light pulse in a cold Λ-type atomic medium.

**II. Physical model and light-atom equations**

We study stationary light pulse interactions in a three-level atomic system composed of two lower long-lived levels, |1> and |2> and one optically excited level |3> (see Fig.1). The transition |1> ↔ |3> is resonant to the input weak signal light pulse while the transition |2> ↔ |3> is driven by two counterpropagating laser fields. The laser fields are characterized by the Rabi frequencies $\Omega_+$, $\Omega_-$ and wavevectors $K_+$, $K_-$ with frequency detuning $\Delta = \omega_c - \omega_{32}$, $\omega_{mn}$ – frequency atomic transition |m> ↔ |n>, $\omega_p$ and $\omega_c$ are the frequency of probe and control fields and $\gamma_2$, $\gamma_3$ are decay rate of the atomic coherences associated with levels |2> and |3>, respectively.

For generation of SLP one can propose various scenarios of external coherent laser control. We briefly describe the temporal scheme analyzed recently in [20]. Here, the weak probe signal field $\hat{E}_+$ is launched into the atomic medium in a presence of one copropagating control laser field $\Omega_+$. We characterize the probe field $\hat{E}_+$ by its wave vector $\bar{k}_+$ and carrier frequency $\omega_+$ with spectral detuning from the optical transition $\Delta = \omega_p - \omega_{31}$. It is also assumed that initially all three-level atoms are prepared on the ground level |1>. After complete arrival of slow light pulse into the medium we can switch off the control laser field following the well-known procedure [3] of EIT based quantum storage on the long-lived atomic coherence of |2> ↔ |3>. Later we *adiabatically*

switch on two counterpropagating control laser fields. Such switching will produce some band gap for the light fields in frequency domain around $\omega_{31}$ [24], at the same time the stored long-lived atomic coherence will be transformed into the optical coherence generating resonant light fields near the band gap spectral range. Below we describe the light-atom dynamics of the excited light field after transient stage of its generation [23], i.e. SLP dynamics.

In order to clarify the basic properties of stationary and quasi-stationary light fields $\hat{E}_{\pm} = \sqrt{\hbar \omega_p/(2\varepsilon_0 V)} \hat{\mathcal{E}}_{\pm}(t,z) e^{-i\omega_p(t \mp \frac{z}{c})} + H.c.$ (where $\hbar$ is the Planck's constant, $\varepsilon_0$ is the electric permittivity, V is a quantization volume) we use coupled Maxwell–Bloch equations for slowly varied atomic coherences $\hat{P}_{12}^j$ and $\hat{P}_{13}^j$ and light field amplitudes $\hat{\mathcal{E}}_{\pm}(t,z)$:

$$\frac{\partial}{\partial t} \hat{P}_{13}^j(t, z_j) = -\gamma_3 \hat{P}_{31}^j(t, z_j) + ig\{\hat{\mathcal{E}}_+(t, z_j)e^{-i(\Delta t - kz_j)} + \hat{\mathcal{E}}_-(t, z_j)e^{-i(\Delta t + kz_j)}\} +$$

$$+i\{\Omega_+ e^{-i(\Delta t - Kz_j)} + \Omega_- e^{-i(\Delta t + Kz_j)}\} \hat{P}_{12}^j(t, z_j), \tag{1}$$

$$\frac{\partial}{\partial t} \hat{P}_{12}^j(t, z_j) = -\gamma_2 \hat{P}_{12}^j(t, z_j) + i\{\Omega_+^* e^{i(\Delta t - Kz_j)} + \Omega_-^* e^{i(\Delta t + Kz_j)}\} \hat{P}_{13}^j(t, z_j), \tag{2}$$

$$\left(\frac{\partial}{c\partial t} - ik_o + \frac{\partial}{\partial z}\right) \hat{\mathcal{E}}_+(t, z) = \frac{iNg}{c} \hat{P}_{13}(t, z) e^{i(\Delta t - kz)}, \tag{3}$$

$$\left(\frac{\partial}{c\partial t} - ik_o - \frac{\partial}{\partial z}\right) \hat{\mathcal{E}}_-(t, z) = \frac{iNg}{c} \hat{P}_{13}(t, z) e^{i(\Delta t + kz)}, \tag{4}$$

where $g = \wp_{13}\sqrt{\frac{\omega_+}{2\varepsilon_0 \hbar V}}$ – constant of atom-photon interaction [2], N is an atomic concentration, $\wp_{13}$ – dipole moment on the atomic transition |1>-|3>, $k_0 = \frac{\omega_{21}}{c}$, $k = \frac{\omega_p}{c}$, $K = \frac{\omega_c}{c}$, $\omega_c$ is a carrier frequency of control laser fields, $\gamma_2$ and $\gamma_3$ are relaxation constants of atomic coherences on the atomic transitions |1>-|2> and |1>-|3>.

By using Fourier transformation $\widetilde{\hat{P}_{nm}}(\omega, z) = \frac{1}{\sqrt{2\pi}} \int_{-\infty}^{\infty} \hat{P}_{nm}(t,z) e^{-i\omega t} dt$, $\widetilde{\hat{\mathcal{E}}}_{\pm}(\omega + \Delta, z) = \frac{1}{\sqrt{2\pi}} \int_{-\infty}^{\infty} \hat{\mathcal{E}}_{\pm}(t,z) e^{-i\Delta t} e^{-i\omega t} dt$, from Eq. (2) we get

$$\widetilde{\hat{P}}_{13}(\omega, z) = \frac{ig(i\omega + \gamma_2)\{\widetilde{\hat{\mathcal{E}}}_+(\omega + \Delta, z) e^{i(kz_j)} + \widetilde{\hat{\mathcal{E}}}_-(\omega + \Delta, z) e^{-i(kz_j)}\}}{(i\omega + \gamma_3)(i\omega + \gamma_2) + \{\Omega_+^2 + \Omega_-^2 + 2\Omega_+ \Omega_- \cos(2Kz)\}}, \tag{5}$$

and after Fourier decomposition of the denominator in (6) we get:

$$\widetilde{P}_{13}(\omega,z) = \left\{\widetilde{\mathcal{E}}_+(\omega+\Delta,z)e^{i(k_+z_j)} + \widetilde{\mathcal{E}}_-(\omega+\Delta,z)e^{-i(k_-z_j)}\right\}\frac{g(i\omega+\gamma_2)}{\sqrt{\Gamma(\omega)^2+(2\Omega_+\Omega_-)^2}} \cdot$$

$$\cdot\left[1 + \sum_{n=1}^{\infty} i^n \left(e^{in(2Kz)} + e^{-in(2Kz)}\right)\frac{(2\Omega_+\Omega_-)^n}{\left(\Gamma(\omega)+\sqrt{\Gamma(\omega)^2+(2\Omega_+\Omega_-)^2}\right)^n}\right], \quad (6)$$

where $\Gamma(\omega) = \omega(\gamma_3+\gamma_2) - i(\gamma_3\gamma_2 - \omega^2 + \Omega_+^2 + \Omega_-^2)$.

Similar expression in the adiabatic limit ($|\omega|\to 0$) have been used in [14,16] for analysis of SLP. Below we use Eq. (6) for analysis of stationary and quasi-stationary light pulses without the adiabatic approximation. By performing spatial Fourier transformation $\widetilde{\widetilde{A}}_\pm(\omega,\tilde{k}) = \frac{1}{\sqrt{2\pi}}\int_{-\infty}^{\infty}\widetilde{A}_\pm(\omega,z)e^{i\tilde{k}z}dz$, we get from Eqs. (3),(4) in the two mode approximation

$$\left(\frac{\omega}{c} - k_0 - \tilde{k}\right)\widetilde{\widetilde{\mathcal{E}}}_+(\omega,\tilde{k}) = \alpha\widetilde{\widetilde{\mathcal{E}}}_+(\omega,\tilde{k}) + \beta\widetilde{\widetilde{\mathcal{E}}}_-(\omega,\tilde{k}), \quad (7)$$

$$\left(\frac{\omega}{c} - k_0 + \tilde{k}\right)\widetilde{\widetilde{\mathcal{E}}}_-(\omega,\tilde{k}) = \alpha\widetilde{\widetilde{\mathcal{E}}}_-(\omega,\tilde{k}) + \beta\widetilde{\widetilde{\mathcal{E}}}_+(\omega,\tilde{k}), \quad (8)$$

where $\alpha = \frac{Ng^2}{c}\frac{(i(\omega-\Delta)+\gamma_2)}{\sqrt{\Gamma(\omega-\Delta)^2+(2\Omega_+\Omega_-)^2}}$, $\beta = i\rho\alpha$, $\rho = \frac{(2\Omega_+\Omega_-)}{\left(\Gamma(\omega-\Delta)+\sqrt{\Gamma(\omega-\Delta)^2+(2\Omega_+\Omega_-)^2}\right)}$. It is worth noting a good acceptability of the two wave approximation has been examined in [24] for analysis of photonic band gap.

Eqs. (7), (8) is characterized by the following dispersion relation

$$\tilde{k}_\pm(\omega) = \pm\sqrt{[\alpha(\omega)+k_0-\omega/c]^2 - \beta^2(\omega)} = \pm K(\omega), \quad (9)$$

with appropriate eigen-states of two light field modes $A_\pm(\omega,\tilde{k})$ of Eqs. (7),(8):

$$A_+(\omega,\tilde{k};z) = \frac{\beta(\omega)}{\frac{\omega}{c}-\alpha(\omega)-k_0-K(\omega)}A(\omega,q)e^{ik_+(\omega)z} + A(\omega,q)e^{-ik_-(\omega)z}, \quad (10)$$

$$A_-(\omega,\tilde{k};z) = \mathcal{B}(\omega,q)e^{-ik_+(\omega)z} + \frac{\beta(\omega)}{\frac{\omega}{c}-\alpha(\omega)-k_0+K(\omega)}\mathcal{B}(\omega,q)e^{ik_-(\omega)z}, \quad (11)$$

where $k_+(\omega) = k + K(\omega)$ and $k_-(\omega) = k - K(\omega)$. As seen in Eqs. (7),(8), the parameter $\rho(\omega)$ and $\beta(\omega)$, respectively, determines the coupling strength of two counterpropagating weak light fields related its eigen-states. The whole SLP field will be in a superposition: $E(\omega,z) = \alpha_+A_+(\omega,\tilde{k};z) +$

$\alpha_- A_-(\omega, \tilde{k}; z)$ with some amplitudes $\alpha_+(\omega)$ and $\alpha_-(\omega)$ determined by the transient stage of SLP generation and initial condition.

We note a quite simple analytical formula in Eq. (9) that makes an easy general spectral analysis for both SLP and usual slow light (when $\Omega_- = 0$ or $\Omega_+ = 0$), where Eq. (9) transforms into well-known EIT dispersion [2]. By using Eq. (9) we can also find analytically the group velocity of the coupled light fields $v_{gr}(\omega) = \frac{\partial \omega}{\partial K}$ for arbitrary strength of the control laser fields $\Omega_-$ and $\Omega_+$. By neglecting small misphasematching ($k_0 \to 0$ [25]) and weak atomic decoherence $\gamma_2 \to 0$ we find very strong coupling regime for $\rho(\omega) > 1$ (see details later). In the opposite case of weak coupling $|\beta^2(\omega)| \ll [\alpha(\omega) + k_0 - \omega/c]^2$ (i.e., for $|\Omega_-| \ll |\Omega_+|$, or $|\Omega_+| \ll |\Omega_-|$) we get almost free uncoupled two light modes: $A_+(\omega, \tilde{k}; z) \approx \frac{\frac{\omega}{c} - \alpha(\omega) - k_0}{2\beta(\omega)} A(\omega, q) e^{ik_+(\omega)z}$ and $A_-(\omega, \tilde{k}; z) \approx \mathcal{B}(\omega, q) e^{-ik_+(\omega)z}$, which are counterpropagating each other along z-axis. More detailed analysis of the spectral and temporal properties of SLP and quasi-stationary light pulse is a subject of the next sections.

### III Properties of stationary light pulse

For a symmetric case of standing control laser field $\Omega_+ = \Omega_- = \Omega$ we get a quite simple formula for the wave number $K(\omega)$ from Eq. (9):

$$K(\omega) = \left(\frac{\omega}{c} + \frac{N\wp^2}{c} \frac{(\omega - i\gamma_2)}{\sqrt{\tilde{\Gamma}_0(\omega)^2 - 4\Omega^4}}\right) \sqrt{1 - \left(\frac{2\Omega^2}{\tilde{\Gamma}_0(\omega) + \sqrt{\tilde{\Gamma}_0(\omega)^2 - 4\Omega^4}}\right)^2}, \qquad (12)$$

where $\tilde{\Gamma}_0(\omega) = [(\gamma_2 + i\omega)(\gamma_3 + i\omega) + 2\Omega^2]$.

Real (dispersion) and Imaginary (absorption) parts of Eq. (12) are depicted in Fig.2.

It is worth noting that spectral behavior of Eq. (12) presented in Fig.2 is in a perfect agreement with numerical results of [21]. For comparison we present the dispersion and absorption Figs. 3 and 4 for stationary with usual traveling control light fields. As shown in Fig.3, dispersion for standing control field is characterized by more steep spectral behavior at line center. Absorption for the standing control field (see Fig.4) demonstrates remarkably linear character with a frequency detuning $|\omega|$ while there is quadratic dependence on the usual EIT absorption. These spectral behaviors of dispersion and absorption explain faster spreading and decay of SLP. From the basic physical point of view it becomes clear if take into account in usual EIT that the effect of control light field characterized by different amplitudes resulting in spreading of dispersion, group velocity and spectral domain of weak absorption. Such spectral properties remain the EIT scheme for surface plasmons [34], where the intensive control plasmon field is also tightly varied in space.

Using Eq. (12) for negligibly weak relaxation $\gamma_2 \to 0$ we get $K(\omega) \approx \frac{\omega^{3/4}}{2v_0} \sqrt[4]{\Omega^2/\gamma_3} e^{-i\frac{\pi}{8}}$ that determines the following spectral group velocity:

$$v_{gr}(\omega) = \frac{\partial \omega}{\partial k} = \frac{8}{3} \sqrt[4]{\frac{\gamma_3}{\Omega}} v_0 \sqrt[4]{\frac{\omega}{\Omega}} \left(\cos\frac{\pi}{8} + i \sin\frac{\pi}{8}\right), \qquad (13)$$

where $v_0 = c \frac{\Omega^2}{N\wp^2}$ is a group velocity in usual EIT.

In accordance with Eq. (13), we get that SLP can be generated only in narrow spectral range ($\omega/\Omega \ll 1$) where $v_{gr}(\omega) \ll v_o \to 0$ (see Fig.5). Imaginary part in Eq. (13) is related to the light field absorption as it is depicted in Fig.4.

For comparison we present a spectral behavior of the group velocities for the standing and traveling control laser fields in Fig.6. It is clearly seen from Figs. 5 and 6 that standing control field can provide SLP in much more narrower spectral range of frequencies in comparison with the standard EIT scheme. These spectral properties clearly explain a previous result [21] that effective generation of SLP is possible only for large enough spatial length providing sufficiently narrow spectral width.

Spectral properties of group velocity and absorption for different amplitudes of standing control laser field are presented in Figs. 7 and 8, where $v_{gr} \sim \Omega^{3/2} |\omega|^{1/4}$ at line center. In Fig. 8 linear spectral dependence of absorption is preserved for different amplitudes of standing control laser field. Here, the absorption decreases as standing control laser field increases. This is consistent with average Autler-Townes splitting of excited optical level due to the driving standing light field. As a result we conclude that, with intense controlling stand light fields, stationary light is generated in more favorable spectral conditions. In this case we provide lower absorption with a small group velocity where $v_{gr}(\omega) \approx 0$. However, due to a very sharp spectral growth of the group velocity of the stationary light non-zero average group velocity results in as shown in Fig.9: one SLP mode $A_+(t,z) = \int d\omega f(\omega) A_+(\omega, \tilde{k}; z) e^{-i\omega t}$ with Gaussian spectral shape $f(\omega) \sim \exp[-\omega^2/2\sigma^2]$. Figure 10 demonstrates dispersion and absorption effects on the SLP mode for the spectral width σ. All three $A_+(t,z)$-SLP modes in Fig. 10 move in +z direction and wider initial spectral width of SLP leads to higher absorption and larger average group velocity. Similar result occurs for $A_-(t,z)$-SLP modes moving in "-z" direction (not shown). So taking into account a symmetry decomposition of the initial SLP on two field modes: $E(t=0,z) = \frac{1}{\sqrt{2}}\{A_+(t=0;z) + A_-(t=0;z)\}$ we get that initial light pulse envelope splits into two pulses propagating in opposite directions that is in a general agreement with adiabatic limit [14,16], and the SLP splitting is accelerated for wider spectral width of the initial light pulse.

Let us also analyze an influence of atomic decoherence $\gamma_2$ between levels 1 and 2 on the group velocity of $A_+(t,z)$ SLP mode. Putting the small spectral detuning $|\omega| \ll \gamma_3$ in Eq. (12) we get

$$K(\omega) = \frac{N g^2}{\Omega^{3/2}} \frac{\omega - i\gamma_2}{2c} \frac{1}{\sqrt[4]{(\gamma_2 + i\omega)\gamma_3}} \approx \frac{N g^2}{2c \sqrt[4]{\gamma_2 \gamma_3} \Omega^{3/2}} \left( \frac{3}{4} \omega - i\gamma_2 (1 + \frac{3}{32} \left(\frac{\omega}{\gamma_2}\right)^2 \right), \tag{14}$$

Where Eq. (14) leads to nonzero group velocity (see also Fig.11) even for central frequency domain $|\omega| < \gamma_2$:

$$v_{gr}(|\omega| < \gamma_2) = \frac{1}{K'(0)} = \frac{8c \sqrt[4]{\gamma_2 \gamma_3} \Omega^{3/2}}{3 N g^2} \sim \sqrt[4]{\gamma_2} > 0. \tag{15}$$

Thus, the atomic decoherence ($\gamma_2 \neq 0$) makes it completely impossible to form a SLP modes for spectral region $\delta\omega_f < \gamma_2$, where the atomic relaxation decouples the counterpropagating modes $\widetilde{\widetilde{\mathcal{E}}}_+(\omega, \tilde{k})$ and $\widetilde{\widetilde{\mathcal{E}}}_-(\omega, \tilde{k})$, so that the forward and backward parts $A_+(t,z)$ and $A_-(t,z)$ of the light pulse can freely move in opposite directions.

**IV Long-lived quasi-stationary light pulse**

As shown in Figs. 5 -7 and 11, the standing control light field drastically changes the spectral properties of dispersion and absorption due to spatial dependence on the control laser field amplitude. Taking into account such influence it seems promising to find some new regimes for such spatially inhomogeneous coherent control of slowdown light fields. Below we propose such a scheme for quasi-stationary light field.

For control counterpropagating laser fields with different amplitudes ($\Omega_- < \Omega_+$) and weak atomic decoherence $\gamma_2 \ll |\Omega_+ - \Omega_-|$ we get from Eq, (9) in the limit of negligibly small spectral detuning $\omega \to 0$ : $v_{gr}(\omega)|_{\omega \to 0} = \frac{1}{K'(0)} = \frac{c}{Ng^2} \Omega_+ (\Omega_+^2 - \Omega_-^2)^{\frac{1}{2}} = v_+ \sqrt{1 - \Omega_-^2/\Omega_+^2} = v_+ \sqrt{1 - v_-^2/v_+^2}$ (where $v_\pm = \frac{c\Omega_\pm^2}{Ng^2}$ is an usual group velocity in EIR with one travelling control light field $\Omega_+$ or $\Omega_-$) that coincides with previous result obtained in the adiabatic limit [14] (see also Refs. [16] and [18]). However it is important to control the group velocity of SLP in wider spectral range comparable with total resonant linewidth $\gamma_3$. Using Eq. (13), we have numerical analyses for the spectral behavior of group velocity as the function of ratio $\Omega_-/\Omega_+$ in Fig.12, where an interesting example of the quasi-stationary light at $\Omega_-/\Omega_+ = 0.66$ is presented. From Fig. 12, we see that the quasi-stationary light can be characterized by flat spectral properties of the group velocity providing thereby weaker spatial spreading of slow light field. The flat behavior is a result of two opposite effects – of the usual EIT dispersion and the spectral dispersion of group velocity inherent to the light in the system of atoms driven by spatially inhomogeneous control laser field depicted in Fig.11. Figure 13 shows absorption curves related to the quasi-stationary light in comparison with slow light and SLP fields. In Fig. 13, the absorption spectrum of the quasi-stationary light remains still low at line center.

Finally we note that the same group velocity can be maintained by varying the Rabi frequencies of both control laser fields linearly as shown in Fig. 14. For a weak counterpropagating control laser fields $\Omega_- \ll \Omega_+$ the following is obtained from Eq.(6):

$$\widetilde{P}_{13}(\omega, z) \cong \chi(\omega; \Omega_+)\{1 - \zeta Cos(2Kz)\}\left\{\widetilde{\mathcal{E}}_+(\omega + \Delta, z)e^{i(kz_j)} + \widetilde{\mathcal{E}}_-(\omega + \Delta, z)e^{-i(kz_j)}\right\}, \qquad (16)$$

where $\chi(\omega; \Omega_+) = \frac{ig(i\omega+\gamma_2)}{\{(i\omega+\gamma_3)(i\omega+\gamma_2)+\{\Omega_+^2+\Omega_-^2\}}$ is related to usual dressed susceptibility of EIT medium in a presence of $\Omega_+$ control field [2], parameter $\zeta = \frac{2\Omega_+\Omega_-}{\{(i\omega+\gamma_3)(i\omega+\gamma_2)+\{\Omega_+^2+\Omega_-^2)\}}$ determines a modulation depth and coupling between the two coupled light fields.

For the large control field $\Omega_+$ and quite narrow bandwidth of the signal light field $|\delta\omega| \ll \Omega_+$, we get $\zeta \cong \frac{2\Omega_-}{\Omega_+}$. This corresponds to the coupling of two slow light fields $\hat{\mathcal{E}}_+(t, z)$ and $\hat{\mathcal{E}}_-(t, z)$ travelling at the EIT condition determined by $\sqrt{\Omega_+^2 + \Omega_-^2} \approx \Omega_+$ control field. Thus the effective formation of quasi-stationary light pulse can be realized by adiabatic switching on the second control field $\Omega_-(t)$ up to $\Omega_- \leq 0.66\Omega_+$ or to some another magnitude determined by the spectral width of the input light field. After such switching the slowdown group velocity will be also determined by the effective coupling of two interacting fields $\hat{\mathcal{E}}_+(t, z)$ and $\hat{\mathcal{E}}_-(t, z)$ in accordance with $v_+\sqrt{1 - \Omega_-^2/\Omega_+^2}$ as discussed above.

**V Conclusion**

We analyzed stationary and quasi-stationary light pulses in a Λ-type cold atomic medium. By developing a model based on Maxwell–Bloch equations for slowly varied amplitudes [14-16,18] we derived quite a simple analytical expression for the dispersion relation of stationary and quasi-stationary light field modes. By using the dispersion relation we performed a detailed analysis of basic spectral properties of SLP and quasi-stationary light pulses. Herein, we revealed quite a narrow spectral range of SLP and observed a strong influence of the atomic decoherence on the spatial splitting of SLP. We also predicted a promising new regime for the generation of quasi-stationary light pulses characterized by weaker spatial spreading in the presence of the control laser fields with appropriate relative intensities. Overall, the performed analysis provided clearer physical picture of the basic properties of SLP in a cold atomic medium, and the proposed scheme of quasi-stationary light pulse control for the application of enhanced nonlinear interactions of weak light pulses at the EIT condition.

Authors thank Russian Foundation for Basic Research through grant no. 12-02-91700 and the Korean government via international cooperation program of NRF-2012K1A2B2A07033421.

**Captions for the figures:**

Fig. 1. A schematic of atomic levels for resonant interactions with weak probe fields $E_p^{\pm}$ and two control fields $\Omega_c^{\pm}$.

Fig. 2. Dispersion – $(c/\gamma 3)Re(K(\omega))$ (red solid line) and absorption $(c/\gamma 3)Im(K(\omega))$ (green dashed line). The curves are presented in $\gamma_3$: $\Omega_+ = \Omega_- = \gamma_3$, $\gamma_2 \to 0$, $\Delta \to 0$.

Fig. 3. EIT dispersion $(c/\gamma_3)Re(K(\omega))$ for travelling (red line $\Omega_+ \neq 0$, $\Omega_- = 0$) and standing (blue line $\Omega_+ = \Omega_-$) control light fields. Other parameters are the same as in Fig. 2: $\Omega_+ = \Omega_- = \gamma_3$, $\gamma_2 \to 0$, $\Delta \to 0$, $\gamma_3 = 1$.

Fig. 4. EIT absorption $(c/\gamma_3)Im(K(\omega))$ for travelling (red line $\Omega_+ \neq 0$, $\Omega_- = 0$) and standing (blue line $\Omega_+ = \Omega_-$) control light fields, other parameters are the same as in Fig.2: $\Omega_+ = \Omega_- = \gamma_3$, $\gamma_2 \to 0$, $\Delta \to 0$, $\gamma_3 = 1$.

Fig. 5. Spectral properties of group velocity for standing control laser field $\Omega_+ = \Omega_-$, other parameters are the same as in Figs. 2-4: $\Omega_\pm = \gamma_3$, $\gamma_2 \to 0$, $\Delta \to 0$, $\gamma_3 = 1$.

Fig. 6. Group velocities for the standing ($\Omega_- = \Omega_+$ red curve) and traveling ($\Omega_- = 0$, $\Omega_+ \neq 0$ blue curve) control laser fields in cold atomic systems. Other parameters are the same as in Figs. 2-4: $\Omega_\pm = \gamma_3$, $\gamma_2 \to 0$, $\Delta \to 0$, $\gamma_3 = 1$.

Fig. 7. Group velocities for different amplitudes of standing control laser field. Other parameters are the same as in Figs. 2-4: $\Omega_\pm = \gamma_3$, $\gamma_2 \to 0$, $\Delta \to 0$, $\gamma_3 = 1$.

Fig. 8. Spectral properties of the absorption for different amplitudes of standing control laser field. Other parameters are the same as in Figs. 2-4: $\Omega_\pm = \gamma_3$, $\gamma_2 \to 0$, $\Delta \to 0$, $\gamma_3 = 1$.

Fig. 9. Spatial profile of SLP mode $A_+(t,z)$ for different moments of time t=0, t=0.2, t=0.4. Time is denoted by $1/\gamma_3$; Spectral width $\sigma = \gamma_3$.

Fig. 10. Spatial profile of SLP mode $A_+(t,z)$ for t=0 and t=0.4 for three spectral widths $\sigma = 1/\sqrt{2}$, 1, 1/2 (in unit of $\gamma_3$).

Fig. 11. Spectral properties of group velocity $v_{gr}(\omega)$ for SLP mode $A_+(t,z)$ at different relaxation rate $\gamma_2 = 0$; 0.02; 0.04 (in unit of $\gamma_3$); larger atomic relaxation exclude any formation of SLP even in narrow spectral range around $\omega = 0$, other parameters are the same as in Figs. 2-4: $\Omega_\pm = \gamma_3$, $\Delta \to 0$.

Fig. 12. Spectral properties of group velocity in particular cases: SLP, $\Omega_- = \Omega_+$ (red curve); usual slow light $\Omega_- = 0$, $\Omega_+ \neq 0$ (blue line); quasi-stationary light $\Omega_- = 0.66\Omega_+$ (green line).

Fig. 13. Spectral properties of absorption $(c/\gamma 3)Im(K(\omega))$ for SLP field $\Omega_- = \Omega_+$ (red triangle curve), usual EIT slow light $\Omega_- = 0, \Omega_+ \neq 0$ (blue curve) and quasi-stationary llight $\Omega_-/\Omega_+ = 0.66$ (green line).

Fig. 14. Relation between the two control Rabi frequencies $\Omega_-$ and $\Omega_+$ providing the same group velocity of quasi-stationary light pulse.

**Fig.1.**

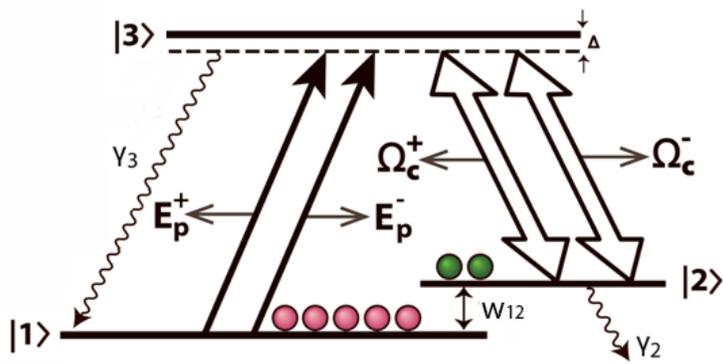

**Fig.2.**

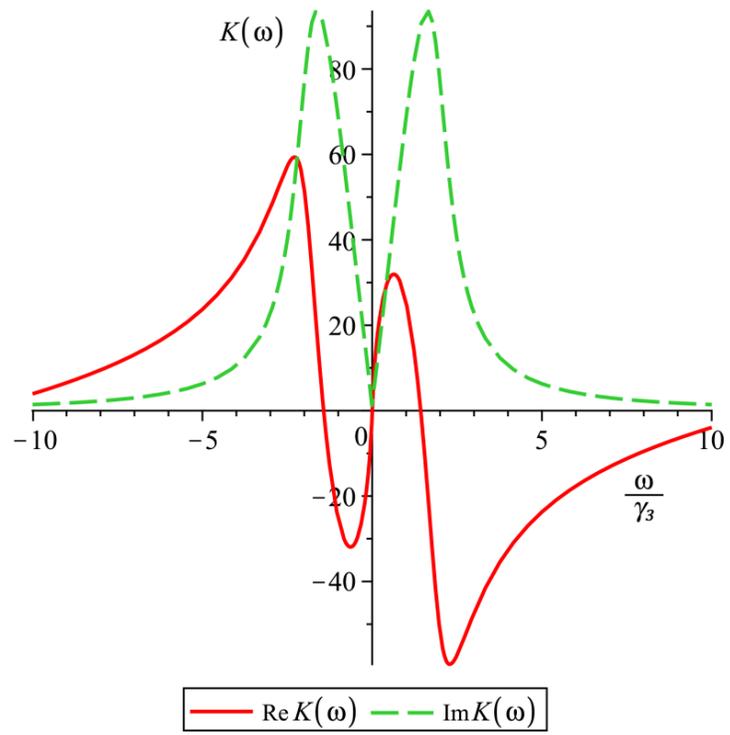

**Fig. 3.**

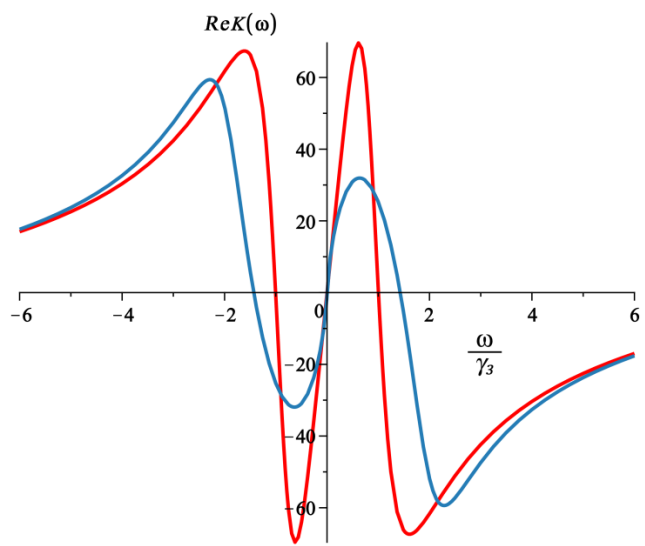

**Fig.4.**

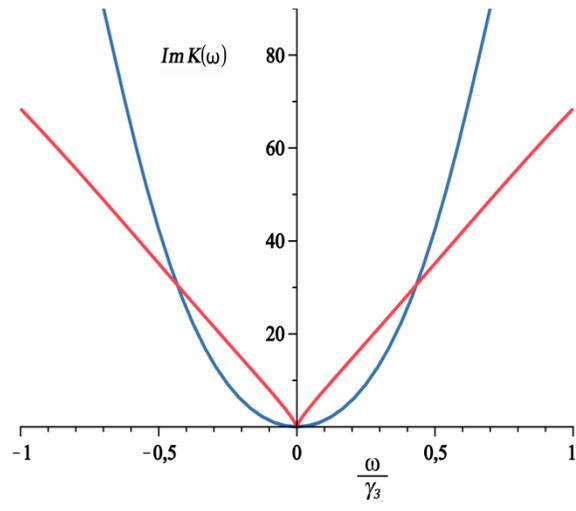

**Fig. 5.**

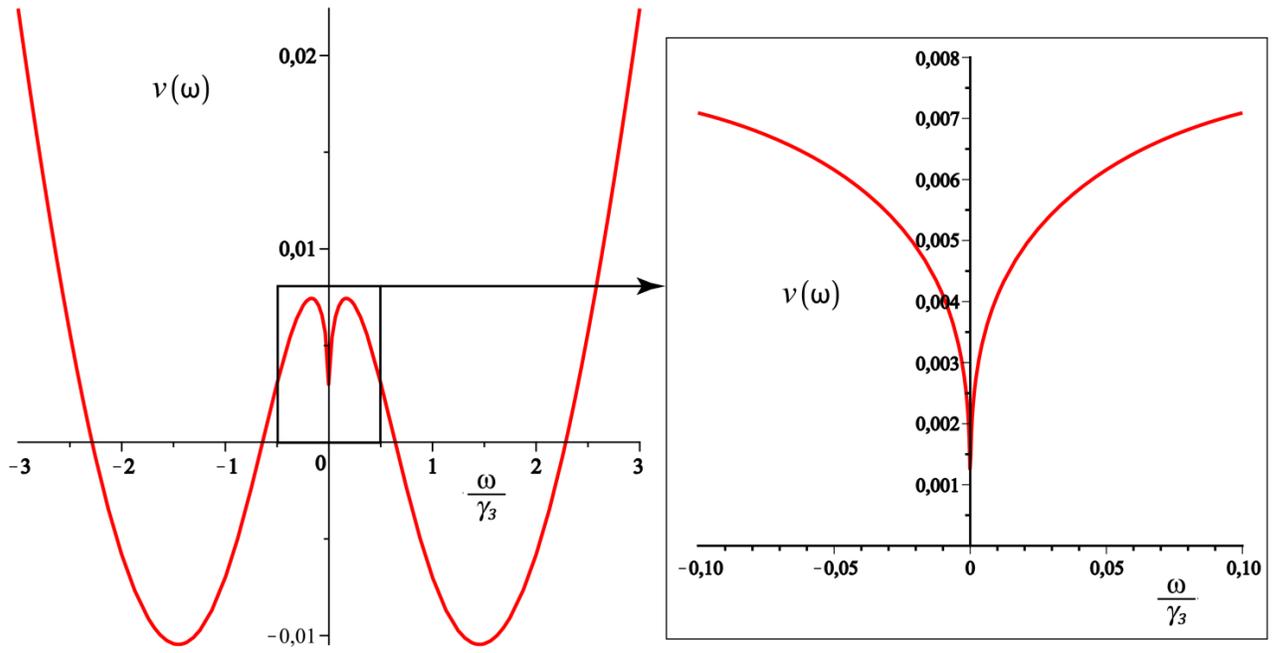

**Fig.6.**

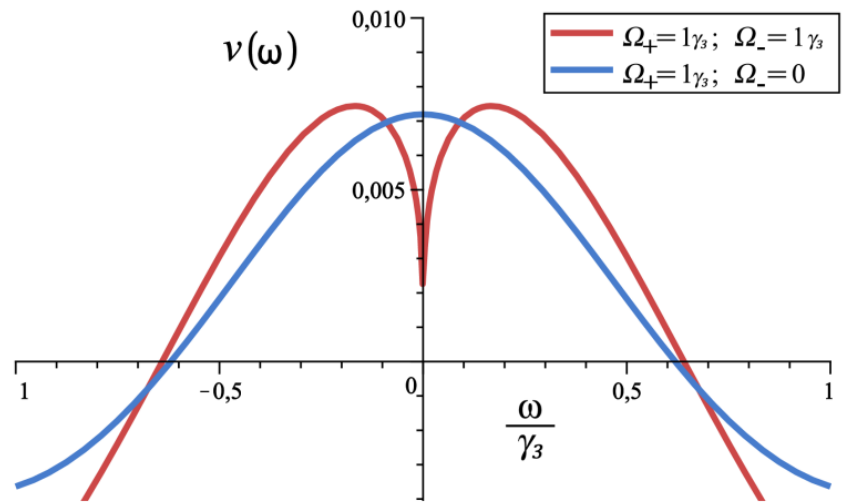

**Fig. 7.**

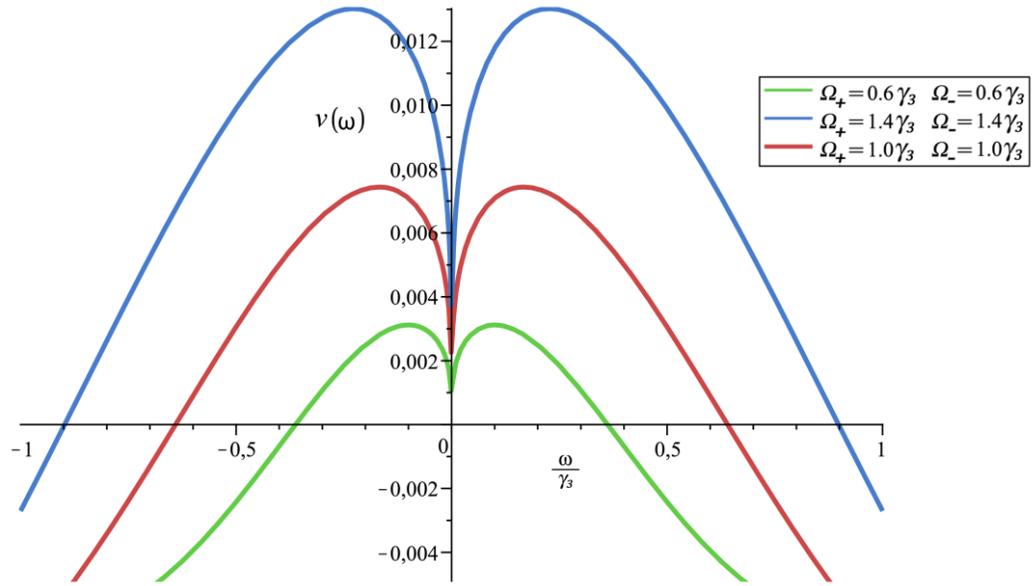

**Fig. 8.**

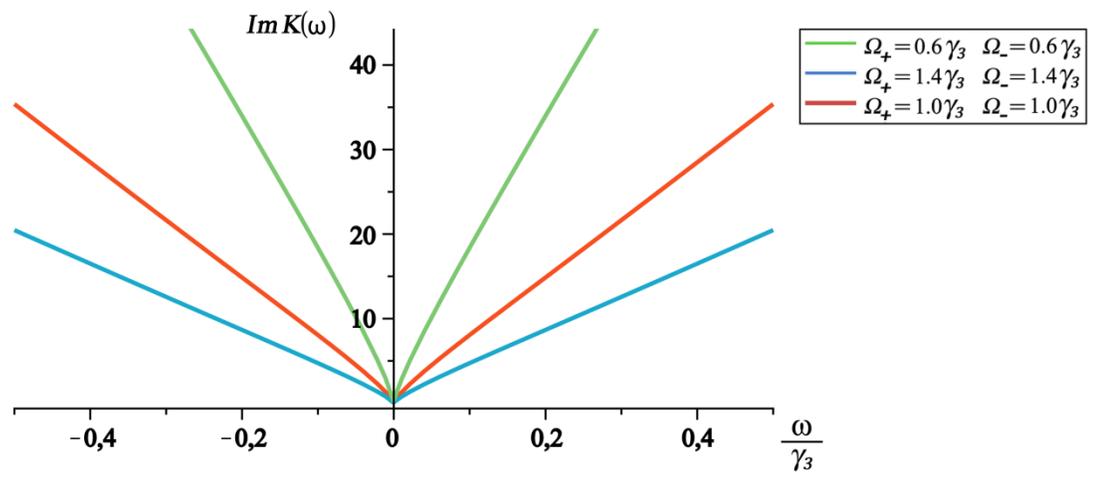

**Fig.9**

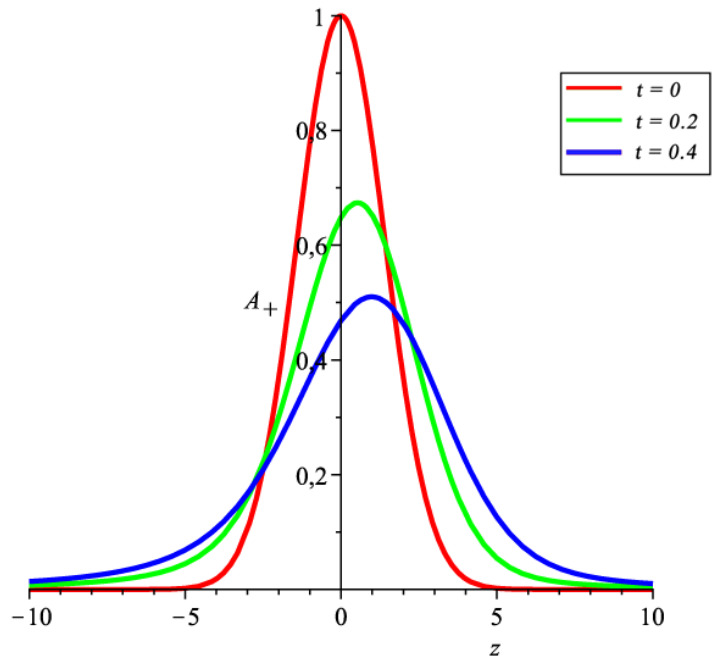

**Fig. 10.**

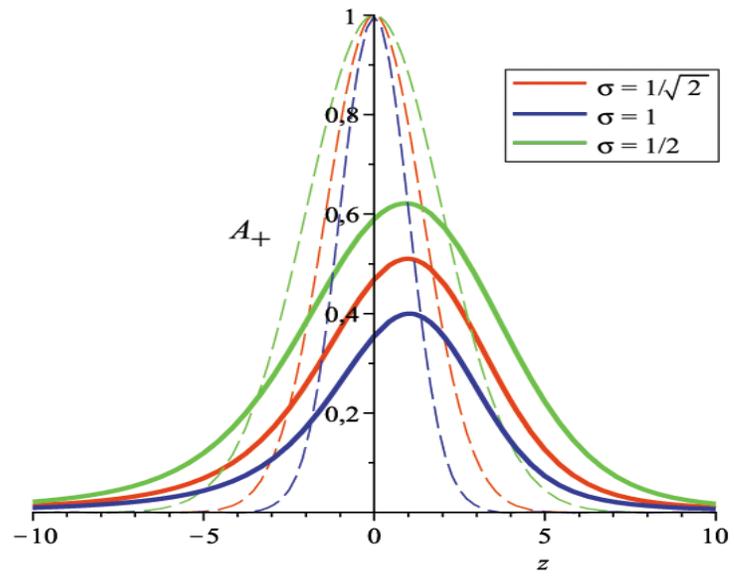

**Fig. 11.**

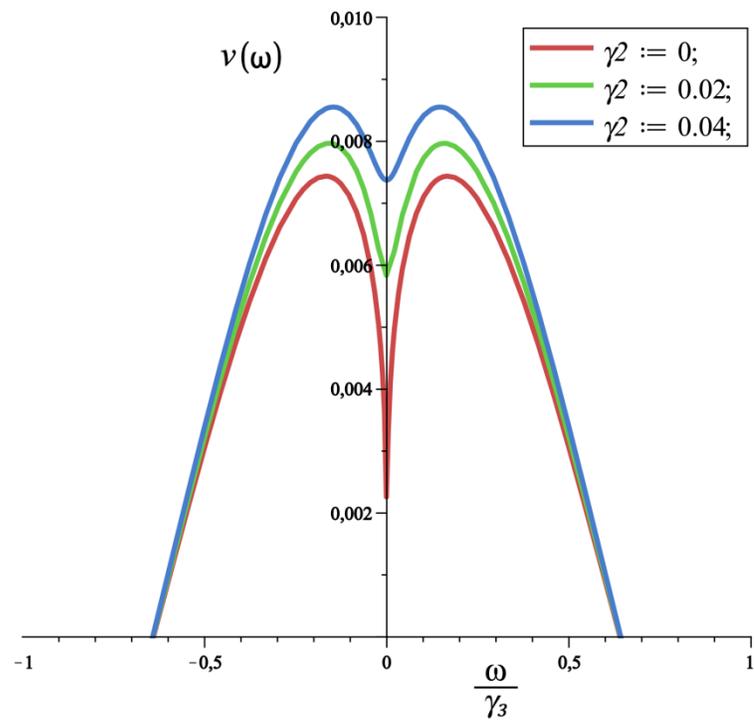

**Fig. 12.**

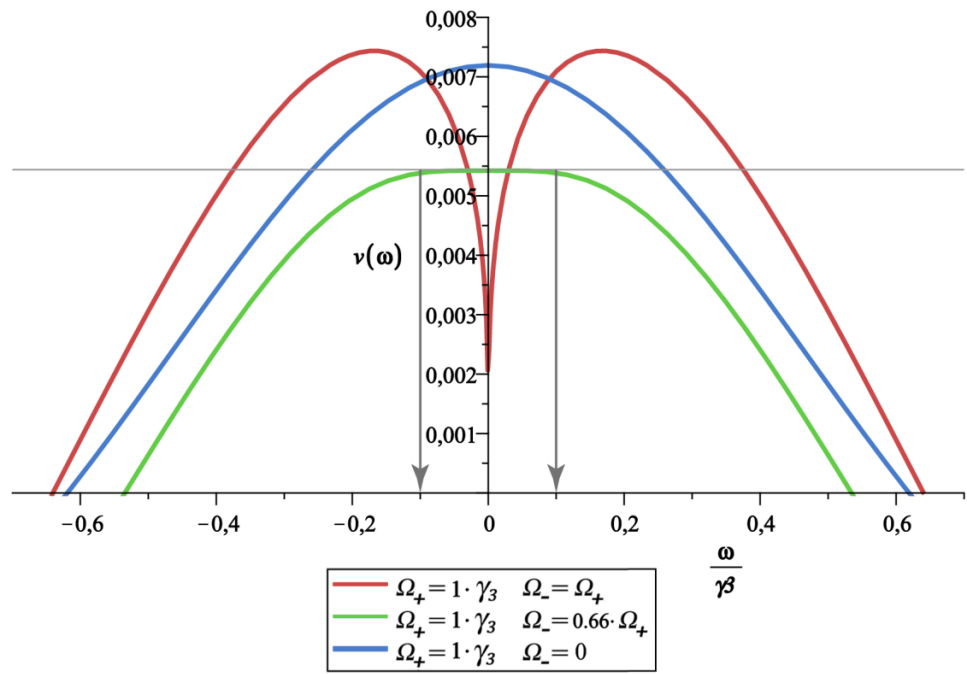

**Fig. 13.**

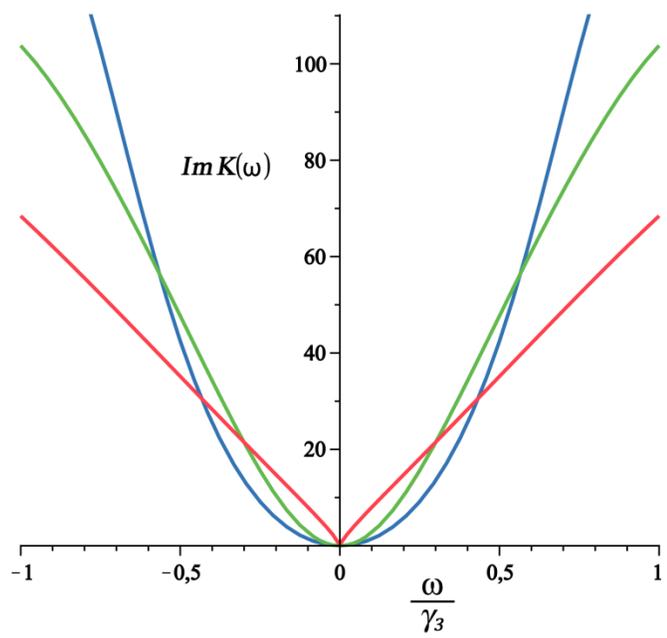

**Fig.14**

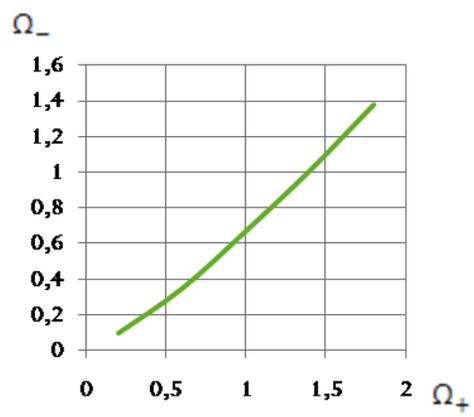